\documentclass[twocolumn,english,aps,prl,showpacs]{revtex4}
\usepackage{amsmath,graphicx,amssymb,epsfig,babel,dsfont}

\renewcommand{\Re}{{\rm Re}}

\newcommand{\Tr}{{\rm Tr}}
\newcommand{\rd}{{\rm d}}

\newcommand{\rs}{{\rm s}}
\newcommand{\rp}{{\rm p}}

\begin{document}

\title{True thermal antenna with hyperbolic metamaterials}

\author{Gr\'egory Barbillon}
\affiliation{Laboratoire Charles Fabry,UMR 8501, Institut d'Optique, CNRS, Universit\'{e} Paris-Sud 11,
2, Avenue Augustin Fresnel, 91127 Palaiseau Cedex, France.}
\author{Jean-Paul Hugonin}
\affiliation{Laboratoire Charles Fabry,UMR 8501, Institut d'Optique, CNRS, Universit\'{e} Paris-Sud 11,
2, Avenue Augustin Fresnel, 91127 Palaiseau Cedex, France.}
\author{Svend-Age Biehs}
\affiliation{Institut f\"{u}r Physik, Carl von Ossietzky Universit\"{a}t, D-26111 Oldenburg, Germany.}
\author{Philippe Ben-Abdallah}
\email{pba@institutoptique.fr}
\affiliation{Laboratoire Charles Fabry,UMR 8501, Institut d'Optique, CNRS, Universit\'{e} Paris-Sud 11,
2, Avenue Augustin Fresnel, 91127 Palaiseau Cedex, France}
\affiliation{Universit\'{e} de Sherbrooke, Department of Mechanical Engineering, Sherbrooke, PQ J1K 2R1, Canada.}

\date{\today}


\begin{abstract}
A thermal antenna is an electromagnetic source which emits in its surrounding, a spatially coherent field  in the infrared frequency range. Usually, its emission pattern changes with the wavelength so that the  heat flux it radiates is weakly directive. Here, we show that a class of hyperbolic materials, possesses a  Brewster angle  which is weakly dependent on the wavelength,  so that they can radiate like a true thermal antenna with a highly directional heat flux. The realization of these sources could  open a new avenue in the field of thermal management  in far-field regime. 
\end{abstract}

\maketitle

The thermal radiation \cite{Planck,Kirchoff} a hot body emits  in its background results from a well-known incoherent emission process. The local charges in the medium (electrons, ions or partial atomic charges) oscillate thanks to thermal fluctuations and as the corresponding oscillators are usually delta correlated they radiate incoherently in their surrounding. A direct consequence of this mechanism is the absence of privileged directions of emission.
However, in 1986 and 1988,  Hesketh et al. \cite{Hesketh86,Hesketh88} showed that a textured surface of a doped silicon sample could behave as a thermal antenna, that is as a spatially coherent source thanks to the presence of a surface plasmon polariton, a surface wave whose the  electromagnetic field is spatially correlated. Since this pioneer work, numerous spatially coherent sources \cite{Kreiter,Greffet,Kollyukh,PBA_JOSA,Celanovic,Lee,Drevillon,Battula,Joulain,Zhang,Wang} have been proposed.
However, sources with extremely directional emission patterns have been achieved so far at a given single frequency mainly. The emission angle of these sources generally changes significantly with respect to the wavelength throughout the Planck window. It follows that the  heat flux they radiate, which results from the spectral integration of the directional monochromatic emissivity weighted by the Planck distribution function, is not notably directional. Today, the development of broadband angular selective sources in the infrared range remains a challenging problem.
In this Letter we show that a class of hyperbolic materials (HM) can be used to achieve a 'true thermal antenna' which radiates a highly directional heat flux in its surrounding.

To start, let us consider an arbitrary semi-infinite planar anisotropic medium at temperature $T$ surrounded by a bosonic field at zero temperature. According to the theory of fluctuational electrodynamics~\cite{RytovBook1989} the  radiative heat flux lost  in its surrounding by this  medium in the direction $\boldsymbol{u}=\frac{c}{\omega}(\boldsymbol{\kappa},\gamma_0)$ with $\gamma_0=\sqrt{\omega^2/c^2-\kappa^2}$ can be formally written  into a Landauer-like form~\cite{PBA2010,Biehs2010,BiehsPRB2016}
\begin{equation}
  \rd\Phi(\boldsymbol{u}) =2\int_{0}^{\infty}\!\!\frac{\rd\omega}{2\pi}\,\Theta(\omega,T)  \mathcal{T}(\omega, \boldsymbol{\kappa}) \frac{\rd^2\boldsymbol{\kappa}}{(2\pi)^2}  , \label{Eq:flux_inf}
\end{equation}
where  $\boldsymbol{\kappa} := (k_x, k_y)^t$  is the parallel component of wavector (with the constraint $|\boldsymbol{\kappa}|<\omega/c$ for the propaging waves), $ \Theta(\omega,T):=\frac{\hbar\omega}{e^{\hbar\omega \beta}- 1}$ is the mean energy of a harmonic oscillator in thermal equilibrium at temperature $T$ and  $ \mathcal{T}$ denotes the transmission coefficient associated to each mode $(\omega, \boldsymbol{\kappa})$ which reads 
\begin{equation}
  \mathcal{T}(\omega, \boldsymbol{\kappa}, d) := \frac{1}{2}\Tr\bigl[(\mathds{1}-\mathds{R}^\dagger \mathds{R})  \bigr].
\label{Eq:TransmissionCoeff}
\end{equation}
Here we have introduced the reflection 
operator for both polarization states ($\rs,\rp$) 
\begin{align}
  \mathds{R}&:= \left(\begin{array}{cc}
r{}^{ss} & r{}^{sp}\\
r{}^{ps} & r{}^{pp}
\end{array}\right).
\end{align}
When further introducing $I^0_\omega(T):= \Theta(\omega,T) \frac{\omega^2}{4\pi^2 c^2}$, the spectral intensity of a blackbody at the frequency $\omega$ and using the generalized emissivity 
\begin{equation}
\begin{split}
  \epsilon(\omega, \boldsymbol{u}) &:= \mathcal{T}(\omega, \boldsymbol{\kappa}) \\
                                   &=\frac{1}{2}\bigl\{ 2- | r^{ss}|^2-| r^{pp}|^2 \\
                                   &\qquad\quad - | r^{sp} |^2-| r^{ps}|^2 \bigr\}, 
\label{Eq:emissivity}
\end{split}
\end{equation}
the  flux radiated by the source in the direction $\boldsymbol{u}$ can be written as
\begin{equation}
 \rd\Phi(\boldsymbol{u}) = \int_{0}^{\infty} \!\! \rd\omega\, \frac{c^2}{\omega^2}\epsilon(\omega, \boldsymbol{u}) I^0_\omega(T) \frac{\rd^2\boldsymbol{\kappa}}{\pi}  .  \label{Eq:power_Planck}
\end{equation}
Notice that expression (\ref{Eq:emissivity}) includes the emissivity in the two different polarization states $s$ and $p$ and the emissivity in the cross-polarized states $sp$ and $ps$ as well allowing so to deal with arbitrary anisotropic sources. When the source displays an azimuthal symmetry the directional heat flux simplifies to
\begin{equation}
 \rd\Phi(\theta) = 2\cos\theta\sin\theta \rd\theta \biggl( \int_{0}^{\infty}\!\! \rd\omega\, \epsilon(\omega, \theta) I^0_\omega(T) \biggr).
\label{Eq:power_Planck2}
\end{equation}
Both expressions (\ref{Eq:power_Planck}) and (\ref{Eq:power_Planck2}) are the classical expressions predicted by the Kirchoff's theory~\cite{Kirchoff}. An inspection of 
this expressions clearly shows that an angular drift in the emission pattern for the source with respect to the frequency leads to a heat flux which is less directive. 
Therefore, in order to get a highly directional heat flux, we need to have a source with a spatial degree of coherence which does not change significantly 
all over the Planck window. 
\begin{figure}[Hhbt]
\includegraphics[scale=0.35]{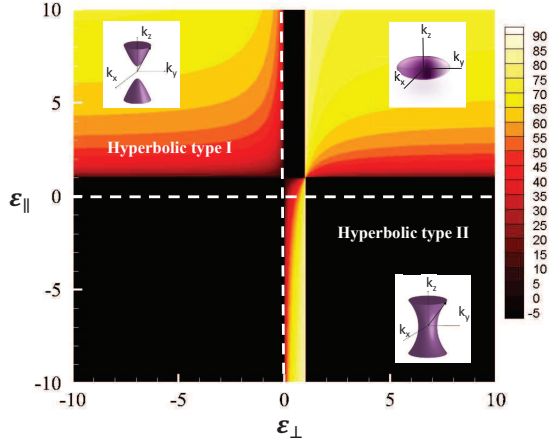}
\caption{ Brewster angle (in degree) in the $(\epsilon_{\perp},\epsilon_{||})$ plane for transparent uniaxial media. The black zones correspond to media which have no Brewster angle. The insets show the iso-frequency surfaces of different uniaxial crystals.  
\label{Brewster}}
\end{figure}

Below, we show that  a class of HM precisely behaves like that. 
To this end, we consider a uniaxial anisotropic medium with a dielectric permittivity tensor $\boldsymbol{\underline{\epsilon}}=\epsilon_{||}(\boldsymbol{x}\otimes\boldsymbol{x}+\boldsymbol{y}\otimes\boldsymbol{y})+\epsilon_{\perp}\boldsymbol{z}\otimes\boldsymbol{z}$ , $\epsilon_{||}$ being the permittivity parallel to the surface and $\epsilon_{\perp}$  the permittivity along the normal to its surface. For this medium the components of the reflection operator are
\begin{align}
r{}^{ss} =\frac{\gamma_0-\gamma_s}{\gamma_0+\gamma_s},\label{reflection1}\\
r{}^{pp}=\frac{\epsilon_{||}\gamma_0-\gamma_p}{\epsilon_{||}\gamma_0+\gamma_p},\label{reflection2}\\
r{}^{ps}=r{}^{sp}=0\label{reflection3}
\end{align}
with $\gamma_s=\sqrt{\epsilon_{||}\omega^2/c^2-\kappa^2}$ and $\gamma_p=\sqrt{\epsilon_{||}\omega^2/c^2-\frac{\epsilon_{||}}{\epsilon_{\perp}}\kappa^2}$. From these expressions 
it can be directly seen that by imposing the condition $r^{pp} = 0$ that the Brewster angle $\theta_B$ is given, using the identity $\kappa=\omega/c\sin\theta_B$, by
\begin{equation}
\theta_B=\arcsin\sqrt{\frac{\epsilon_{\perp}(\epsilon_{||}-1)}{\epsilon_{||}\epsilon_{\perp}-1}}.   \label{Eq:Brewster2}
\end{equation}
\begin{figure}[Hhbt]
\includegraphics[scale=0.35]{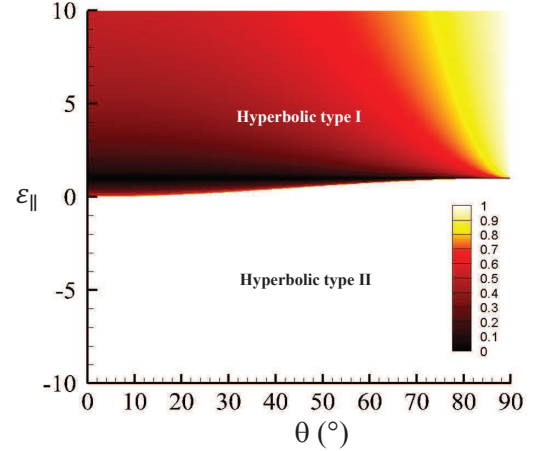}
\caption{ Reflection in polarization s of HM for the type I $(\epsilon_{||}>0)$ and type II $(\epsilon_{||}<0)$. The light zone corresponds to the region where the reflection is close to 1.  
\label{reflection_r_ss}}
\end{figure}
In Fig.1 this angle is plotted  in the $(\epsilon_{||},\epsilon_{\perp})$ plane that is for  arbitrary uniaxial media when losses are negligible. Each quadrant corresponds to a specific class of anisotropic medium. When both parameters $\epsilon_{||}$ and $\epsilon_{\perp}$ are positive the medium is a standard uniaxial crystal with an ellipsoidal iso-frequency surface. On the contrary, if both parameters are negative the medium is a metallic-like anisotropic medium. In this case, the iso-frequency surface is purely imaginary and the medium does not support propagating modes. In the two others quadrants, $\epsilon_{||}$ and $\epsilon_{\perp}$ have opposite signs. In both cases, the iso-frequency relations $\frac{\kappa_x^2+\kappa_y^2}{\epsilon_{\perp}}+\frac{\gamma_z^2}{\epsilon_{||}}=\frac{\omega^2}{c^2}$ define hyberboloidal surfaces. When $\epsilon_{||}>0$ and $\epsilon_{\perp}<0$  the HM is called type I HM while in the case where $\epsilon_{||}<0$ and $\epsilon_{\perp}>0$ it is of type II. One associates to these two types of HM two different iso-frequency surfaces: a one two-sheeted hyperboloid for type I HM and a one-sheeted hyperboloid for type II HM. As shown in Fig.~1 both types of HM possesses a Brewster angle in some regions of $(\epsilon_{||},\epsilon_{\perp})$ plane. For the type I HMs, this angle only exists when $\epsilon_{||} \leq 1$ and we observe its drift towards grazing angles as the value of $\epsilon_{||}$ increases. On the contrary for type II HMs, a Brewster angle only exists when $\epsilon_{\perp} \leq 1$. But what is worthwhile to note is that the Brewster angle changes very few with the value of $\epsilon_{||}$ making these media potentially good candidates to design a coherent thermal antenna with a weak angular variation of emission angle with respect to the frequency.

\begin{figure}[Hhbt]
\includegraphics[scale=0.35]{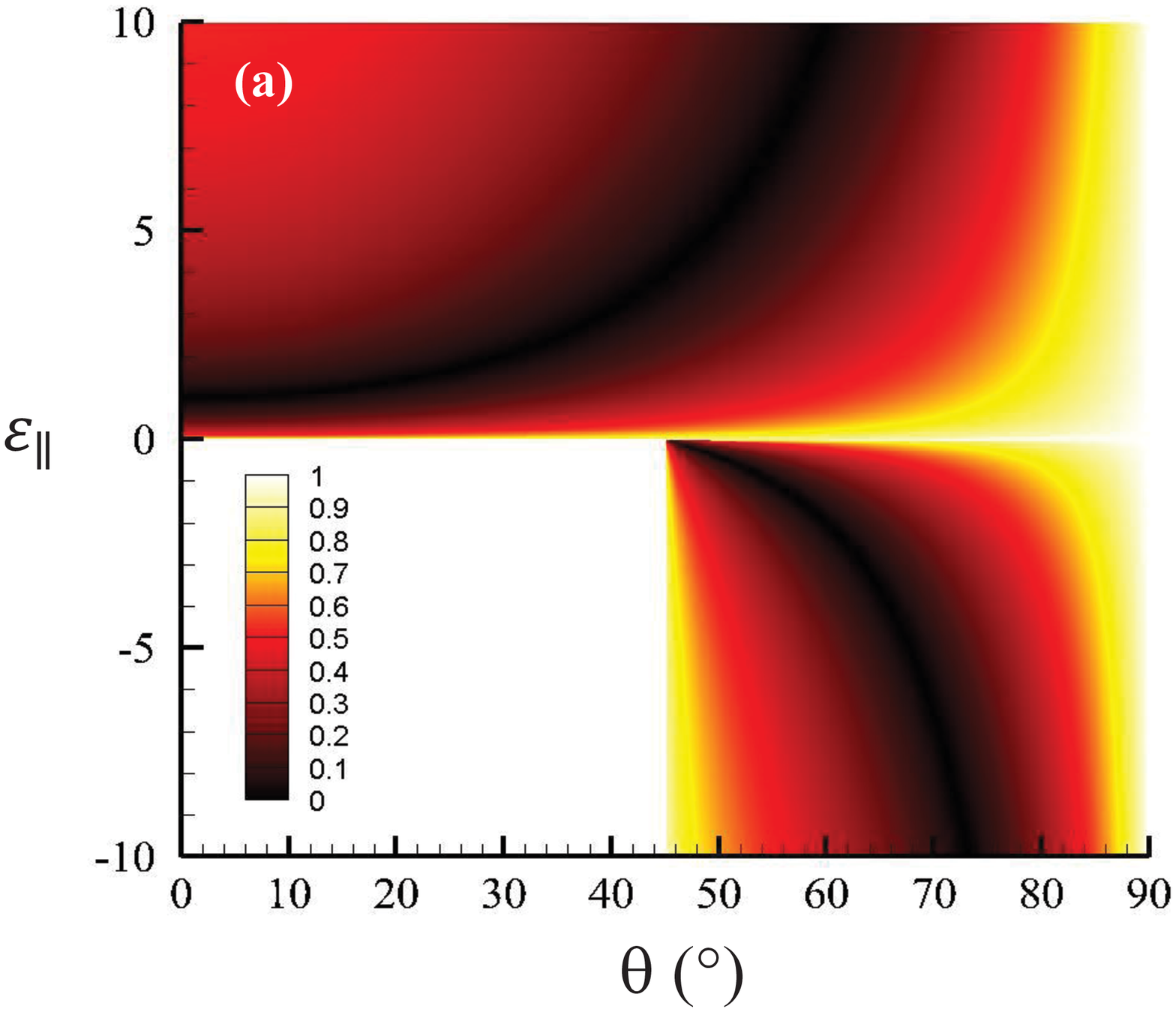}
\includegraphics[scale=0.35]{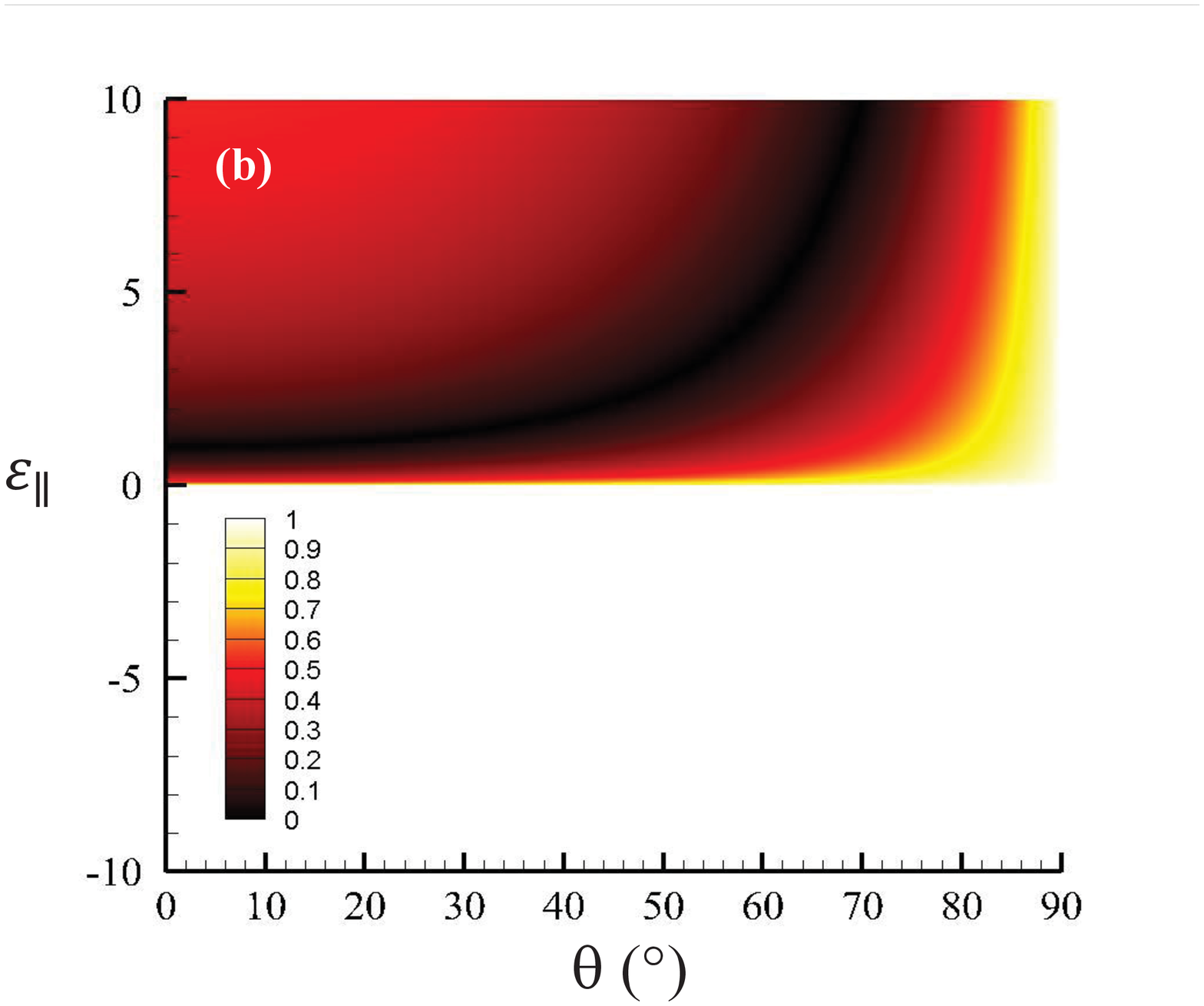}
\caption{ Reflection in polarization p of HM for type I $(\epsilon_{||}>0)$ and type II $(\epsilon_{||}<0)$  when (a) $(\mid\epsilon_{\perp}\mid=0.5)$ and (b) $(\mid\epsilon_{\perp}\mid=5)$ . 
\label{reflection_r_pp}}
\end{figure}

To confirm this prediction, let us investigate the reflection coefficients of HMs.  
These reflections coefficients are plotted  in Figs.~2 and 3. In s-polarization, we see that, while the reflection of type I HMs is relatively weak for any angle of incidence, the reflection of type II HMs is very close to 1 showing so that their thermal emission is almost entirely p-polarized. Moreover, the reflection coefficient in p-polarization of type I and II HMs, plotted in Fig.~3 for two different values of $\epsilon_{\perp}$, confirm the weak variation of Brewster angle of HMs of  type II when $\epsilon_{\perp}$ is smaller than one. They also demonstrate, according to relation (\ref{Eq:emissivity}) that the thermal emission of these media is restricted to an angular sector beyond a critical angle. For $\epsilon_{\perp}=0.5$ (see Fig.3-a) this angle is located around $45^\circ$ . This critical angle increases (not shown in Fig.~3) with the value of $\epsilon_{\perp}$. It can be derived by the condition that $\gamma_p = 0$ which is equivalent to $\kappa = \epsilon_\perp \omega/c$ or $\theta = \arcsin(\epsilon_\perp)$~\cite{HuChui2002} showing again that the condition $0 \leq \epsilon_\perp \leq 1$ must be necessarily fulfilled.

Finding a natural HM of type II that displayes the required properties over a broad spectral range in the Planck frequency window is a tricky task. 
However, a metamaterial can be designed for that purpose. To this end, we consider an artificial composite structure
formed by alternating layers of materials of permittivity $\epsilon_1$ and  $\epsilon_2$. In the long-wavelength limit 
the structure behaves like an homogeneous uniaxial crystal~\cite{YehBook} with 
\begin{eqnarray}
\epsilon_{||}&=&f\epsilon_{\rm 1} +(1- f)\epsilon_{\rm 2}, \label{Eq:eps_par}\\
\epsilon_{\perp}&=&\frac{\epsilon_{\rm 1}\epsilon_2}{f\epsilon_{\rm 2}+(1-f)\epsilon_1},\label{Eq:eps_perp}
\end{eqnarray}
where $f$ denotes the filling factor with respect to medium 1. It is easy to show that to obtain a HM of type II with $\epsilon_{\perp}<1$ the two dielectric constants must satisfy the following inequalities
\begin{eqnarray}
f\epsilon_{1}+(1-f)\epsilon_{2}<0, \label{Ineq1}\\
\epsilon_{2}(\epsilon_{\rm 1}-f)+(f-1)\epsilon_{\rm 1}<0, \label{Ineq2}\\
f\epsilon_{2}+(1-f)\epsilon_{\rm 1}<0,\label{Ineq3}\\
\epsilon_{1}\epsilon_{2}<0.\label{Ineq4}
\end{eqnarray}
A graphical solution of this system (see  Supplemental Material~\cite{SupplMat}) shows that the possible values of $\epsilon_1$ and $\epsilon_2$ are restricted to a special
region which itself depends on the filling fraction. On the other hand, when fixing $\epsilon_1$ and $\epsilon_2$ this system of conditions leads to limits of the filling fraction
for which we can expect to have a broad band Brewster angle. Guided by these conditions we consider as a realistic example of a hyperbolic thermal antenna a metamaterial 
composed by alternating Zirconium Nitride (ZrN) and gold (Au) layers. ZrN is a low index dielectric in the infrared range ($\Re(\epsilon_1)$ varies between $0.59$ and $0.95$ in the
wavelength range between $4\,\mu{\rm m}$ and $80\,\mu{\rm m}$.~\cite{Palik}).  As for the second material we take gold whose the dielectric constant is well described by a simple Drude model 
\begin{equation}
\epsilon_2=1-\frac{\omega_p^2}{\omega(\omega+i\gamma)},
\end{equation}
with the plasma frequency  $\omega_p = 13.71\times10^{15}\,{\rm rad/s}$  and the electron damping $\gamma = 4.05\times10^{13} {\rm rad/s}$. 

\begin{figure}[Hhbt]
\includegraphics[scale=0.35]{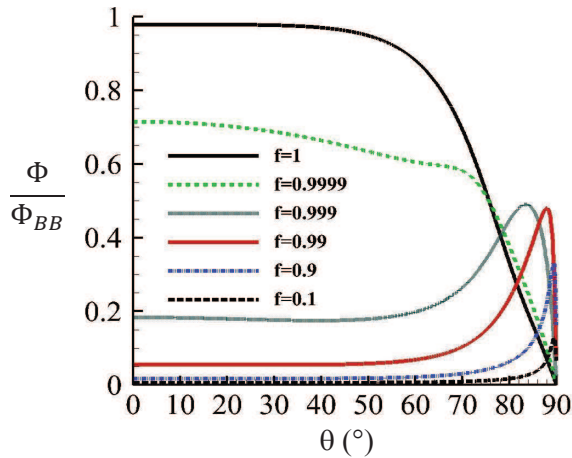}
\caption{ Angular heat flux emitted by a ZrN-Au layered structure at $T=300\,{\rm K}$ for different filling factors in ZrN. The flux is normalized by the blackbody emission.
\label{reflection_r_pp2}}
\end{figure}

In Fig.~4 we plot the angular heat flux at an equilibrium temperature of $300 K$ emitted by ZrN-Au layered structures with different filling factors.  For this temperature, the Wien's wavelength is $\lambda_W=9.66 \mu m$ and about $95\%$ of the radiative energy is emitted  by the source between $0.5\lambda_W=4.83\mu m$ and $4.5\lambda_W=43.47\mu m$. This spectral  range corresponds precisely to  the region where  ZrN has a low dielectric constant. For small filling factors, the structure has a metallic behavior very similar to that one of a gold sample and it is therefore weakly emitting. However, as the filling factor of ZrN increases, a lobe of emission evolves at oblique incidence reaching a maximum value when $f = 0.99$ which means that only $1\%$ of the structure is metallic .
The directivity observed for $f = 0.99$ results from the invariance of Brewster angle.  On the
other hand, since the reflectivity for s-polarization  is almost equal to 1, the thermal emission at the
Brewster angle equals only one-half of the blackbody value. Note that, for other filling fractions like $f = 0.1$ we can also see an emission peak in Fig.~4. From the values of the 
permittivities for $f = 0.1$ we have (see ~\cite{SupplMat}) $\epsilon_\perp > 1$. Hence the emission peak for $f=0.1$ is not due to a pure broadband Brewster angle effect. This conclusion is backed by the fact that the thermal emission is much lower than one-half of the blackbody value. 
 
Although the thermal antenna designed above radiates most of its radiative power in a specific direction of space, its angle of thermal emission is very oblique. However, in principle, according to expression (\ref{Eq:Brewster2}) a thermal antenna can be engineered to emit at an arbitrary 
emission angle by properly tuning the value of $\epsilon_{\perp}$. In particular, we note that when $\epsilon_{\perp}\rightarrow 0$ 
(i.e. for epsilon near zero (ENZ) HM of type II) the emission angle is orthogonal to the surface. 
Around this value, the emission angle varies as  $\theta_B\simeq\sqrt{\epsilon_{\perp}(\mid\epsilon_{||}\mid+1)}$ so 
that provided that $\epsilon_{||}$ is not too large, this angle remains close to zero. 

In summary, we have predicted that a class of HM can emit  most of their radiative power in privilegied directions of space. This result paves the way for highly directive radiative heat sources. We believe that these true thermal antennas should find  broad applications in the field  of fundamental sciences and for a number of applications such as
thermal management, heat-driven control of chemical reaction or thermal regulation in biology.  However, further works are needed to achieve true thermal antenna in  arbitrary emission angles. This will require to find the appropriate effective properties of HMs. Alternatively, another direction of research could be the developement of broadband angular selective  photonic hypercrystals \cite{Narimanov}, metamaterials with periodic spatial variation of both anisotropic and  isotropic materials, which could combine the strong selectivity of photonic crystals with the singular properties of anisotropic media. 
%
%

\begin{acknowledgments}

\end{acknowledgments}


\end{document}